\documentclass[fleqn,a4paper,11pt,notitlepage]{article}
\usepackage[english,german]{babel}
\usepackage{graphics}
\newcommand{\vertcll}{          %
\begin{minipage}{20mm}          %
\unitlength1mm                  %
\begin{picture}(20,6)           %
\thinlines                      %
\put( 0,3){\line(1, 0){20}}     %
\put(10,3){\circle*{3}}         %
\put( 3,3){\line(2, 1){4}}      %
\put( 3,3){\line(2,-1){4}}      %
\put(13,3){\line(2, 1){4}}      %
\put(13,3){\line(2,-1){4}}      %
\end{picture}                   %
\end{minipage}}                 %

\newcommand{\vertclr}{          %
\begin{minipage}{20mm}          %
\unitlength1mm                  %
\begin{picture}(20,6)           %
\thinlines                      %
\put( 0,3){\line( 1, 0){20}}    %
\put(10,3){\circle*{3}}         %
\put( 3,3){\line( 2, 1){4}}     %
\put( 3,3){\line( 2,-1){4}}     %
\put(17,3){\line(-2, 1){4}}     %
\put(17,3){\line(-2,-1){4}}     %
\end{picture}                   %
\end{minipage}}                 %

\newcommand{\vertcrl}{          %
\begin{minipage}{20mm}          %
\unitlength1mm                  %
\begin{picture}(20,6)           %
\thinlines                      %
\put( 0,3){\line( 1, 0){20}}    %
\put(10,3){\circle*{3}}         %
\put( 7,3){\line(-2, 1){4}}     %
\put( 7,3){\line(-2,-1){4}}     %
\put(13,3){\line( 2, 1){4}}     %
\put(13,3){\line( 2,-1){4}}     %
\end{picture}                   %
\end{minipage}}                 %

\newcommand{\vertcrr}{          %
\begin{minipage}{20mm}          %
\unitlength1mm                  %
\begin{picture}(20,6)           %
\thinlines                      %
\put( 0,3){\line( 1, 0){20}}    %
\put(10,3){\circle*{3}}         %
\put( 7,3){\line(-2, 1){4}}     %
\put( 7,3){\line(-2,-1){4}}     %
\put(17,3){\line(-2, 1){4}}     %
\put(17,3){\line(-2,-1){4}}     %
\end{picture}                   %
\end{minipage}}                 %

\newcommand{\vertoll}{          %
\begin{minipage}{20mm}          %
\unitlength1mm                  %
\begin{picture}(20,6)           %
\thinlines                      %
\put( 0,3){\line(1, 0){20}}     %
\put(10,3){\circle{3}}          %
\put( 3,3){\line(2, 1){4}}      %
\put( 3,3){\line(2,-1){4}}      %
\put(13,3){\line(2, 1){4}}      %
\put(13,3){\line(2,-1){4}}      %
\end{picture}                   %
\end{minipage}}                 %

\newcommand{\vertolr}{          %
\begin{minipage}{20mm}          %
\unitlength1mm                  %
\begin{picture}(20,6)           %
\thinlines                      %
\put( 0,3){\line( 1, 0){20}}    %
\put(10,3){\circle{3}}          %
\put( 3,3){\line( 2, 1){4}}     %
\put( 3,3){\line( 2,-1){4}}     %
\put(17,3){\line(-2, 1){4}}     %
\put(17,3){\line(-2,-1){4}}     %
\end{picture}                   %
\end{minipage}}                 %

\newcommand{\vertorl}{          %
\begin{minipage}{20mm}          %
\unitlength1mm                  %
\begin{picture}(20,6)           %
\thinlines                      %
\put( 0,3){\line( 1, 0){20}}    %
\put(10,3){\circle{3}}          %
\put( 7,3){\line(-2, 1){4}}     %
\put( 7,3){\line(-2,-1){4}}     %
\put(13,3){\line( 2, 1){4}}     %
\put(13,3){\line( 2,-1){4}}     %
\end{picture}                   %
\end{minipage}}                 %

\newcommand{\vertorr}{          %
\begin{minipage}{20mm}          %
\unitlength1mm                  %
\begin{picture}(20,6)           %
\thinlines                      %
\put( 0,3){\line( 1, 0){20}}    %
\put(10,3){\circle{3}}          %
\put( 7,3){\line(-2, 1){4}}     %
\put( 7,3){\line(-2,-1){4}}     %
\put(17,3){\line(-2, 1){4}}     %
\put(17,3){\line(-2,-1){4}}     %
\end{picture}                   %
\end{minipage}}                 %

\newcommand{\vvertclra}{        %
\begin{minipage}{30mm}          %
\unitlength1mm                  %
\begin{picture}(30,6)           %
\thinlines                      %
\put( 0,3){\line( 1, 0){30}}    %
\put(10,3){\circle*{3}}         %
\put(20,3){\circle*{3}}         %
\put( 3,3){\line( 2, 1){4}}     %
\put( 3,3){\line( 2,-1){4}}     %
\put(13,3){\line( 2, 1){4}}     %
\put(13,3){\line( 2,-1){4}}     %
\put(27,3){\line(-2, 1){4}}     %
\put(27,3){\line(-2,-1){4}}     %
\end{picture}                   %
\end{minipage}}                 %

\newcommand{\vvertclrb}{        %
\begin{minipage}{30mm}          %
\unitlength1mm                  %
\begin{picture}(30,6)           %
\thinlines                      %
\put( 0,3){\line( 1, 0){30}}    %
\put(10,3){\circle*{3}}         %
\put(20,3){\circle*{3}}         %
\put( 3,3){\line( 2, 1){4}}     %
\put( 3,3){\line( 2,-1){4}}     %
\put(17,3){\line(-2, 1){4}}     %
\put(17,3){\line(-2,-1){4}}     %
\put(27,3){\line(-2, 1){4}}     %
\put(27,3){\line(-2,-1){4}}     %
\end{picture}                   %
\end{minipage}}                 %

\newcommand{\vvertclr}{         %
\begin{minipage}{30mm}          %
\unitlength1mm                  %
\begin{picture}(30,6)           %
\thinlines                      %
\put( 0,3){\line( 1, 0){30}}    %
\put(10,3){\circle*{3}}         %
\put(20,3){\circle*{3}}         %
\put( 3,3){\line( 2, 1){4}}     %
\put( 3,3){\line( 2,-1){4}}     %
\put(27,3){\line(-2, 1){4}}     %
\put(27,3){\line(-2,-1){4}}     %
\end{picture}                   %
\end{minipage}}                 %

\newcommand{\vertring}{         %
\begin{minipage}{40mm}          %
\unitlength1mm                  %
\begin{picture}(40,40)          %
\thinlines                      %
\put( 5, 5){\circle*{3}}        %
\put( 5, 5){\line( 1, 0){15}}   %
\put(20, 5){\circle{3}}         %
\put(20, 5){\line( 1, 0){15}}   %
\put(35, 5){\circle*{3}}        %
\put(35, 5){\line( 0, 1){15}}   %
\put(35,20){\circle{3}}         %
\put(35,20){\line( 0, 1){15}}   %
\put(35,35){\circle*{3}}        %
\put(35,35){\line(-1, 0){15}}   %
\put(20,35){\circle{3}}         %
\put(20,35){\line(-1, 0){15}}   %
\put( 5,35){\circle*{3}}        %
\put( 5,35){\line( 0,-1){15}}   %
\put( 5,20){\circle{3}}         %
\put( 5,20){\line( 0,-1){15}}   %
\end{picture}                   %
\end{minipage}}                 %

\newcommand{\tr}{\mbox{tr }}
\newcommand{\zp}[1]{#1}
\newcommand{\zn}[1]{\overline #1}
\newcommand{\foh}{\ensuremath{\frac{1}{2}}}
\newcommand{\fth}{\ensuremath{\frac{3}{2}}}
\newcommand{\be}{\begin{equation}}
\newcommand{\ee}{\end{equation}}
\newcommand{\bea}{\begin{eqnarray}}
\newcommand{\eea}{\end{eqnarray}}
\newcommand{\beas}{\begin{eqnarray*}}
\newcommand{\eeas}{\end{eqnarray*}}
\newcommand{\pairperm}{\ensuremath{P_{i,j}}}
\newcommand{\pairmagn}{\ensuremath{S_i^z + S_j^z}}
\newcommand{\neel}{N\'{e}el}
\newcommand{\ras}[1]{\renewcommand{\arraystretch}{#1}}
\renewcommand{\phi}{\varphi}
\renewcommand{\rho}{\varrho}
\begin{document}
\selectlanguage{english}
\renewcommand{\thefootnote}{\ensuremath{\fnsymbol{footnote}}}
\begin{titlepage}
\noindent
\LARGE \begin{flushleft}
Optimum ground states for spin-\fth\ chains\footnote{Work
performed within the research program of the Sonderforschungsbereich
341, K\"{o}ln-Aachen-J\"{u}lich}
       \end{flushleft}
\large \begin{flushleft}
H.~Niggemann, J.~Zittartz
       \end{flushleft}
\scriptsize \begin{flushleft}
Institut f\"{u}r Theoretische Physik, Universit\"{a}t zu K\"{o}ln, \\
Z\"{u}lpicher Strasse 77, D-50937 K\"{o}ln, Germany                \\
(email: hn@thp.uni-koeln.de, zitt@thp.uni-koeln.de)
            \end{flushleft}
\large \begin{flushleft}
\today
       \end{flushleft}
\normalsize
\noindent
{\bf Abstract.} We present a set of {\em optimum ground states} for a large class of
spin-\fth\ chains. Such global ground states are simultaneously ground
states of the local Hamiltonian, i.e. the nearest neighbour interaction
in the present case. They are constructed in the form of a matrix product.
We find three types of phases, namely a {\em weak antiferromagnet}, a
{\em weak ferromagnet}, and a {\em dimerized antiferromagnet}.
The main physical properties of these phases are calculated exactly by
using a transfer matrix technique, in particular magnetization and two
spin correlations. Depending on the model parameters, they show a surprisingly
rich structure.
\end{titlepage}
\setcounter{footnote}{0}
\renewcommand{\thefootnote}{\arabic{footnote}}
\newpage
\section{Introduction}
Quantum spin systems have been investigated extensively in recent years,
in particular in one dimension. Experimentally many quasi-one-dimensional
materials have been discovered for quantum spin $S=\foh,1,\fth,2,\ldots$
and interesting physical properties have been observed [1-9]. Most
experiments indicate that such systems have restricted symmetries and that
anisotropy effects are quite important.

Theoretical interest has been mostly initiated by Haldane [10,11] who
emphasized essential differences between integral and half-integral spin
systems which could lead to qualitatively different physical properties.
For integer spin he showed the existence of a new antiferromagnetic phase,
i.e. the {\em Haldane phase}, characterized by a unique ground state, a
gap to the excitations, and exponential decay of correlations. In subsequent
work exact ground states have been constructed for such integer spin models,
mostly for spin 1, and mathematical properties of such states have been
investigated in much detail (see for instace [12-14]).

In recent work [15-18] we have considered a most general class of an\-iso\-tro\-pic
spin-1 chains. Exact ground states were found in a large parameter subspace
of this model in the form of a matrix product and have been called
{\em matrix product ground states}. In the present paper (also \cite{dipl})
we shall apply the same idea, namely the construction of exact ground states
via matrix products, to the case of spin $S=\fth$. It turns out that in this
case three different types of homogeneous states can be constructed, namely
a {\em weak antiferromagnet}, a {\em weak ferromagnet}, and a {\em dimerized
antiferromagnet}, all of them characterized by a broken symmetry, and thus
by long-range order. Of course we find no {\em Haldane antiferromagnet}.

The paper is organized as follows. In section \ref{optgs} we introduce the
notion {\em optimum ground state} for a quite general lattice model whose
global Hamiltonian is the sum of suitably defined local Hamiltonians, for
instance nearest neighbour interactions. It has the property that this global
ground state is simultaneously ground state to all local Hamiltonians.

In section \ref{ham} we consider and classify the Hamiltonian for most
general spin-\fth\ lattice models with nearest neighbour interaction
exhibiting some minimum set of symmetries. The parameter space turns out to
be 12-dimensional.

In section \ref{mpg} it is shown that optimum ground states in the sense of
section \ref{optgs} can also be constructed for spin-\fth\ chains in the
form of matrix products provided certain conditions are satisfied. These
will lead to restrictions on the interaction parameters of the general model
in the following subsections.

Section \ref{gs32chain} contains the main results. In three subsections three
different types of optimum ground states in the form of matrix products are
presented. As mentioned before they describe a {\em weak antiferromagnet},
a {\em weak ferromagnet}, and a {\em dimerized antiferromagnet}. In each case
the single site state matrices and the parameter restrictions are determined.
Also the most important ground state properties are explicitly calculated,
such as magnetizations, two-spin correlations, and corresponding correlation
lengths. All truncated correlations show exponential decay. The main steps
of calculating such properties via transfer matrix technique are briefly
derived in Appendix A.

In section \ref{vertrep} we describe an alternative graphic representation
of optimum ground states as a {\em vertex state model}. This is formulated
for the spin-\fth\ chain, but can be generalized to other spin values $S$ and
to applications in higher dimensions \cite{nkz}. Section \ref{summary} contains
a short summary of main results.
\section{Optimum ground states}
\label{optgs}
We consider a spin system on an arbitrary lattice with homogeneous nearest
neighbour interaction. The Hamiltonian of such a system is of the form
\be
H = \sum_{\langle i,j\rangle} h_{i,j} \quad .
\ee
Angular brackets $\langle i,j\rangle$ denote nearest neighbour lattice
sites $i$ and $j$, to which spin variables are attached.
As the system is homogeneous, all local Hamiltonians
$h_{i,j}$ are in fact equal, they only act on different spin pairs.
By simply adding a trivial constant to the Hamiltonian -- which never
changes the physics of the system -- one can always achieve that the
lowest eigenvalue of $h$ vanishes, i.e. $e_0=0$. In this case, the lowest eigenvalue
of $H$ -- i.~e.~the global ground state energy $E_0$ -- is greater than or equal to zero,
since the sum of non-negative operators is also non-negative. Now one
has to distinguish between two cases:

A) $E_0 > 0$

This is the usual case. The ground state of the complete spin system
involves not only local ground states of $h$ but also (possibly all)
excited states of the local Hamiltonian. As there is no local
condition for the global ground state, one is dealing with a true
many body problem. Therefore the construction of the global ground state of such a
system is usually extremely difficult or even impossible.

B) $E_0 = 0$

In this very special case,
there exists a local condition for finding the ground state. Let
$|\Psi_0\rangle$ be the ground state of $H$. Then we have
\be
0 = \langle\Psi_0|H      |\Psi_0\rangle = \sum_{\langle i,j\rangle}
    \langle\Psi_0|h_{i,j}|\Psi_0\rangle\quad .
\ee
Since all $h_{i,j}$ are non-negative operators, every single
$\langle\Psi_0|h_{i,j}|\Psi_0\rangle$ must vanish. If $\Delta$ is the largest eigenvalue
of $h_{i,j}$, we have the estimate
\be
\langle\Psi_0 | h^2_{i,j} | \Psi_0\rangle
\leq \Delta \cdot \langle\Psi_0 | h_{i,j} | \Psi_0\rangle = 0 \quad .
\ee
This of course requires
\be
\label{optcon}
h_{i,j} |\Psi_0\rangle = 0 \quad,\quad \mbox{for all} \quad \langle i,j\rangle \quad .
\ee
This means: The global ground state consists only of ground states
of the local Hamiltonian, no excited local states are involved. A
ground state of this type is called {\em optimum ground state}. To obtain
optimum ground states for a given system we have to perform two steps.
First, the ground states of the local Hamiltonian must be determined,
which is a local problem with few degrees of freedom. Second, one has to check,
whether these local ground states can be stuck together to form a
global ground state. The most important feature of optimum ground states
is the existence of the local condition (\ref{optcon}). In section
\ref{mpg} we show how optimum ground states can easily be constructed by using
matrix products.
\section{Spin-\fth\ Hamiltonians}
\label{ham}
The physically most interesting systems are those exhibiting a certain
degree of symmetry, as they are most often realized in experimental situations.
We therefore restrict ourselves to Hamiltonians
with only nearest neighbour interaction (for simplicity) and the following set of
symmetries:
\begin{itemize}
\item Homogeneity: All local Hamiltonians $h_{i,j}$ are equal, they only
      act on different spin pairs.
\item Parity invariance: $h_{i,j}$ commutes with the parity operator
      \pairperm, which interchanges the two neighbour spins $i$ and $j$.
\item Rotational invariance in the xy-plane of spin space: $h_{i,j}$
      commutes with the pair magnetization operator \pairmagn.
\item Spin flip invariance: $S^z \rightarrow -S^z$ leaves $h_{i,j}$
      unchanged.
\end{itemize}
In the following, we focus on spin-\fth\ systems. In this case, the
operators $h_{i,j}$, \pairperm\ and \pairmagn\ act on a 16-dimensional
vector space (the space of a spin pair). These operators commute with each
other, so they can be diagonalized simultaneously. It turns out that
\pairperm\ and \pairmagn\ divide the vector space of a spin pair
in small subspaces of dimensions less or equal to~2. Therefore, the
eigenstates of the local Hamiltonian are restricted to not more than
2-dimensional invariant subspaces. This enables us to give a simple
parametrization of all possible local Hamiltonians in terms of projectors
onto the local eigenstates $|v_k\rangle$:
\be
h = \sum_{k=1}^{16} \lambda_k |v_k\rangle\langle v_k| \quad , \quad 
\lambda_k \geq 0 \quad .
\ee
The $\lambda_k$ are real parameters that can be chosen arbitrarily
without breaking one of the above symmetries.

Within a two-dimensional invariant subspace, the two eigenstates
$|v_k\rangle$ of $h$ are not uniquely fixed by the operators \pairperm\ and
\pairmagn. Instead, they form an ortho-normal basis that may be rotated by a certain
angle. Besides the $\lambda$-parameters, one therefore has to introduce one
additional parameter for each two-dimensional invariant subspace in order
to take care of these possible rotations.

A convenient notation for the canonical spin-\fth\ basis is
\ras{1.5}
\be \begin{array}{rclcrcl}
S_i^z\; |\zp{3}\rangle & = & \hphantom{-}\fth |\zp{3}\rangle & \hspace{1cm} &
S_i^z\; |\zp{1}\rangle & = & \hphantom{-}\foh |\zp{1}\rangle \\
S_i^z\; |\zn{3}\rangle & = & -\fth |\zn{3}\rangle & \hspace{1cm} &
S_i^z\; |\zn{1}\rangle & = & -\foh |\zn{1}\rangle \quad .
\end{array} \ee
\ras{1.0}
The following table contains a classification of all
eigenstates\footnote{The states are not normalized, because all normalization
factors can be absorbed into the $\lambda$-parameters.}
of $h$ according to the corresponding eigenvalues $\mu$ and $p$ of the
operators \pairmagn\ and \pairperm, respectively:
\ras{1.5}
\be \begin{array}{llclcl}
\mu= 3,&p= 1 &:&v_3      &=&|\zp{3}\zp{3}\rangle\\
\mu=-3,&p= 1 &:&v_{-3}   &=&|\zn{3}\zn{3}\rangle\\
\mu= 2,&p= 1 &:&v_2^+    &=&|\zp{3}\zp{1}\rangle+|\zp{1}\zp{3}\rangle\\
\mu= 2,&p=-1 &:&v_2^-    &=&|\zp{3}\zp{1}\rangle-|\zp{1}\zp{3}\rangle\\
\mu=-2,&p= 1 &:&v_{-2}^+ &=&|\zn{3}\zn{1}\rangle+|\zn{1}\zn{3}\rangle\\
\mu=-2,&p=-1 &:&v_{-2}^- &=&|\zn{3}\zn{1}\rangle-|\zn{1}\zn{3}\rangle\\
\mu= 1,&p= 1 &:&v_{11}^+ &=&  |\zp{1}\zp{1}\rangle+\frac{a}{2}
                           (\,|\zp{3}\zn{1}\rangle+|\zn{1}\zp{3}\rangle\,)\\
\mu= 1,&p= 1 &:&v_{12}^+ &=&a |\zp{1}\zp{1}\rangle-
                           (\,|\zp{3}\zn{1}\rangle+|\zn{1}\zp{3}\rangle\,)\\
\mu= 1,&p=-1 &:&v_1^-    &=&|\zp{3}\zn{1}\rangle-|\zn{1}\zp{3}\rangle\\
\mu=-1,&p= 1 &:&v_{-11}^+&=&  |\zn{1}\zn{1}\rangle+\frac{a}{2}
                           (\,|\zn{3}\zp{1}\rangle+|\zp{1}\zn{3}\rangle\,)\\
\mu=-1,&p= 1 &:&v_{-12}^+&=&a |\zn{1}\zn{1}\rangle-
                           (\,|\zn{3}\zp{1}\rangle+|\zp{1}\zn{3}\rangle\,)\\
\mu=-1,&p=-1 &:&v_{-1}^- &=&|\zn{3}\zp{1}\rangle-|\zp{1}\zn{3}\rangle\\
\mu= 0,&p= 1 &:&v_{01}^+ &=& (\,|\zp{1}\zn{1}\rangle+|\zn{1}\zp{1}\rangle\,)+
                            b(\,|\zp{3}\zn{3}\rangle+|\zn{3}\zp{3}\rangle\,)\\
\mu= 0,&p= 1 &:&v_{02}^+ &=&b(\,|\zp{1}\zn{1}\rangle+|\zn{1}\zp{1}\rangle\,)-
                             (\,|\zp{3}\zn{3}\rangle+|\zn{3}\zp{3}\rangle\,)\\
\mu= 0,&p=-1 &:&v_{01}^- &=& (\,|\zp{1}\zn{1}\rangle-|\zn{1}\zp{1}\rangle\,)+
                            c(\,|\zp{3}\zn{3}\rangle-|\zn{3}\zp{3}\rangle\,)\\
\mu= 0,&p=-1 &:&v_{02}^- &=&c(\,|\zp{1}\zn{1}\rangle-|\zn{1}\zp{1}\rangle\,)-
                             (\,|\zp{3}\zn{3}\rangle-|\zn{3}\zp{3}\rangle\,) \quad .
   \end{array} \ee
\ras{1.0}
States with $p=1$ are symmetric under spin transposition, the others are
antisymmetric. The parameters $a$, $b$, and $c$ serve to handle the possible
rotations in the two-dimensional invariant subspaces.

Because of spin flip invariance, not all $\lambda$-parameters are
independent. A spin flip transforms an eigenstate with pair magnetization
$\mu$ into a corresponding eigenstate with pair magnetization $-\mu$. These
two eigenstates must have the same energy and therefore the same $\lambda$-parameter.
This leads to the following most general form for all local
spin-\fth\ Hamiltonians which exhibit the required set of symmetries:
\ras{1.5}
\be \begin{array}{rcl}
\label{genham}
h      & = & \lambda_3      (\,|v_3      \rangle\langle v_3      |    + 
                               |v_{-3}   \rangle\langle v_{-3}   |\,) + \\
       &   & \lambda_2^+    (\,|v_2^+    \rangle\langle v_2^+    |    +
                               |v_{-2}^+ \rangle\langle v_{-2}^+ |\,) + \\
       &   & \lambda_2^-    (\,|v_2^-    \rangle\langle v_2^-    |    +
                               |v_{-2}^- \rangle\langle v_{-2}^- |\,) + \\
       &   & \lambda_{11}^+ (\,|v_{11}^+ \rangle\langle v_{11}^+ |    +
                               |v_{-11}^+\rangle\langle v_{-11}^+|\,) + \\
       &   & \lambda_{12}^+ (\,|v_{12}^+ \rangle\langle v_{12}^+ |    +
                               |v_{-12}^+\rangle\langle v_{-12}^+|\,) + \\
       &   & \lambda_1^-    (\,|v_1^-    \rangle\langle v_1^-    |    +
                               |v_{-1}^- \rangle\langle v_{-1}^- |\,) + \\
       &   & \lambda_{01}^+    |v_{01}^+ \rangle\langle v_{01}^+ |    +
             \lambda_{02}^+    |v_{02}^+ \rangle\langle v_{02}^+ |    + \\
       &   & \lambda_{01}^-    |v_{01}^- \rangle\langle v_{01}^- |    +
             \lambda_{02}^-    |v_{02}^- \rangle\langle v_{02}^- | \quad .
\end{array} \ee
\ras{1.0}
The parameter space of all possible local Hamiltonians is 12-dimensional by fixing
the local ground state energy at zero, i.e. $e_0\equiv\lambda_{\mbox{min}}=0$ (as
required in section \ref{optgs}), as we are left with 9 $\lambda$-parameters ($\geq 0$) and
the 3 superposition parameters $a,b,c$. These 12 parameters include a trivial scale
parameter, thus we have 11 non-trivial (interaction) parameters.

We briefly mention the usual representation of the local interaction $h_{i,j}$ in terms
of the canonical spin operators $S^x,S^y,S^z$. Because of the required symmetries, only
polynomials of 3 pair operators are allowed,
\be \begin{array}{rcl}
U & = & S_i^x S_j^x + S_i^y S_j^y \\
V & = & S_i^z S_j^z               \\
W & = & (S_i^z)^2 + (S_j^z)^2 \quad .
    \end{array} \ee
Furthermore, for spin \fth\ at most second powers of spin operators are required, as
higher powers are reduced by the spin algebra. Then it is easily seen that the most
general interaction $h_{i,j}$ can be written as
\be \begin{array}{rcl}
\label{genhamc}
h_{ij} & = & \mbox{const.} + \alpha_1 U + \alpha_2 V + \\
       &   & \alpha_3 U^2 + \alpha_4 (UV+VU) + \alpha_5 V^2 + \alpha_6 W + \\
       &   & \alpha_7 U^3 + \alpha_8 UVU + \alpha_9 VUV + \alpha_{10} V^3 + \\
       &   & \alpha_{11} (UW+WU) + \alpha_{12} (VW+WV) \quad ,
    \end{array} \ee
in terms of 12 interaction parameters $\alpha_n$. Of course, this representation
(\ref{genhamc}) is completely equivalent to (\ref{genham}). We refrain from recalculating
the $\alpha$-parameters in terms of the $\lambda$-parameters and $a,b,c$ and vice versa.
This is simple algebra and is not needed in the following.
\section{Matrix product representation}
\label{mpg}
In one dimension, i.~e.~for spin chains, optimum ground states can easily be
constructed by using matrix products. This method was introduced in [15-17]
and extended in \cite{lkz} to local Hamiltonians with next-nearest neighbour
interactions. The Hamiltonian for a chain of length $L$ with periodic boundary
conditions is
\be
H = \sum_{i=1}^L h_{i,i+1} \quad .
\ee
Now consider a $2\times 2$ matrix with single spin states as its elements
\be
m = \left( \begin{array}{cc}
           |s_1\rangle & |s_2\rangle \\
           |s_3\rangle & |s_4\rangle
           \end{array} \right) \quad .
\ee
The product of two such matrices is defined as
\be
(m\cdot g)_{kl} = \sum_n m_{kn} \otimes g_{nl} \quad ,
\ee
i.~e.~the tensorial product is applied to the elements. Obviously, the trace
of an $L$-fold product of such matrices is a valid $L$-spin state and can
serve as an ansatz for the groundstate of the system. In principle, all these
matrices along the chain can be different, but in our case it is sufficient
to use at most an
alternating product of only two different matrices\footnote{To avoid possible
misfits for closed chains we shall also restrict ourselves to even values of $L$.}
\be
\label{mpgstate}
|\Psi_0\rangle = \tr m_1 \cdot g_2 \cdot m_3 \cdot g_4 \cdots g_L \quad .
\ee
To be an optimum ground state of the spin chain, $|\Psi_0\rangle$ has to be
annihilated by all local Hamiltonians $h_{i,i+1}$. This leads to the following
theorem:
\begin{quote}
If the local Hamiltonian annihilates all elements of the two matrix products
$m\cdot g$ and $g\cdot m$, then $|\Psi_0\rangle$ is an optimum ground state of
the global Hamiltonian $H$.
\end{quote}
\section{Ground states for spin-\fth\ chains}
\label{gs32chain}
In the following subsections we construct three different types of optimum ground states
for spin-\fth\ chains using the matrix product technique. We also calculate the
single spin magnetization and two point correlation functions.
\subsection{Weak antiferromagnet}
\label{wafgs}
An interesting set of ground states is obtained by using the alternating (along the chain)
product of the two matrices
\be
\label{mandg}
   m = \left( \begin{array}{rr}
                 |\zp{1}\rangle &      a |\zp{3}\rangle \\
                 |\zn{1}\rangle & \sigma |\zp{1}\rangle
              \end{array} \right) \quad , \quad
   g = \left( \begin{array}{rr}
          \sigma |\zn{1}\rangle & |\zp{1}\rangle \\
               a |\zn{3}\rangle & |\zn{1}\rangle
              \end{array} \right)
\ee
with the discrete parameter $\sigma=\pm 1$ and an arbitrary real parameter $a$.
The elements of the two matrix products $m\cdot g$ and $g\cdot m$ are the
pair states
\be \begin{array}{rcl}
|\zp{1}\zp{1}\rangle & + & a |\zp{3}\zn{1}\rangle \\
|\zp{1}\zp{1}\rangle & + & a |\zn{1}\zp{3}\rangle \\
|\zn{1}\zn{1}\rangle & + & a |\zn{3}\zp{1}\rangle \\
|\zn{1}\zn{1}\rangle & + & a |\zp{1}\zn{3}\rangle \\
|\zp{1}\zn{1}\rangle & + & \sigma a^2 |\zp{3}\zn{3}\rangle \\
|\zn{1}\zp{1}\rangle & + & \sigma a^2 |\zn{3}\zp{3}\rangle \\
|\zp{1}\zn{1}\rangle & + & \sigma |\zn{1}\zp{1}\rangle \quad .
\end{array} \ee
If $h$ annihilates these 7 pair states, then
\be
\label{wafmpgstate}
|\psi_0\rangle = \tr m_1 \cdot g_2 \cdot m_3 \cdot g_4 \cdots g_L
\ee
is an optimum ground state of the global Hamiltonian $H$. This is
achieved if the following constraints are imposed on the parameters
defined in equation~(\ref{genham}):
\be \begin{array}{l}
\lambda_{11}^+ = \lambda_1^- = \lambda_{01}^\sigma = \lambda_{02}^\sigma =
\lambda_{01}^{-\sigma} = 0 \\
b = c = \sigma a^2 \\
\lambda_3,\;\lambda_2^\sigma,\;\lambda_2^{-\sigma},\;
\lambda_{12}^+,\;\lambda_{02}^{-\sigma} > 0 \\
a \; \mbox{arbitrary,} \; \sigma=\pm 1 \quad .
\end{array} \ee
The inequalities ensure, that all other pair states have a higher energy level.
In this case, the local Hamiltonian $h$ has 7 ground states and 9 excited
states. The system still contains 6 continuous and 1 discrete parameter ($\sigma$).

Expectation values of observables can be calculated using the transfer matrix
technique, which is explained in \cite{ksz1}, for any given chain length $L$. However,
we shall only give the results for the thermodynamic limit $L\to\infty$, since this is
the most important case. The transfer matrix and its eigensystem can be found in
Appendix \ref{appwaftm}. The most interesting expectation value is the single spin
magnetization. If $\langle S_{m/g}^z\rangle_{\psi_0}$ denotes the magnetization
of a spin, represented by an $m$- or $g$-matrix, respectively, we have
\be
\langle S_m^z\rangle_{\psi_0} = -\langle S_g^z\rangle_{\psi_0} =
\frac{1}{2} + \frac{a^2 - 1}{\sqrt{4+(a^2-1)^2}}
\in\left[\frac{1}{2}-\frac{1}{\sqrt{5}},\frac{3}{2}\right[ \quad ,
\ee
which is calculated from the formulae of Appendix A.
So the single spin magnetization alternates from site to site,
which indicates an antiferromagnet with \neel\ ordering. The global magnetization
vanishes. As the sublattices are only fully polarized if $a\to\infty$, we call this
ground state a {\em weak antiferromagnet}.

Unlike the single spin magnetization, the average of the square of the operator $S^z$
covers the full range of possible expectation values:
\ras{1.5}
\be \begin{array}{l}
\langle (S_m^z)^2\rangle_{\psi_0} = \langle (S_g^z)^2\rangle_{\psi_0} \\
=\frac{1}{4} + \frac{4 a^2}{(a^2 - 1)^2 + 4 - (a^2 - 3)
\sqrt{(a^2 - 1)^2 + 4}}\in\left[\frac{1}{4},\frac{9}{4}\right[ \quad .
\end{array} \ee
\ras{1.0}
Because of rotational invariance in the xy-plane, this also gives
\ras{1.5}
\be \begin{array}{l}
\langle (S^x)^2\rangle_{\psi_0} = \langle (S^y)^2\rangle_{\psi_0} =
\foh\left[ \fth\left(\fth + 1\right) - \langle (S^z)^2\rangle_{\psi_0} \right] \\
= \frac{7}{4} - \frac{2 a^2}{(a^2 - 1)^2 + 4 - (a^2 - 3)\sqrt{(a^2 - 1)^2 + 4}}
\quad .
\end{array} \ee
\ras{1.0}

\begin{figure}[htb]
\begin{tabular}{rc}
\raisebox{3cm}{$\langle S_i^z S_{i+1}^z\rangle_{\psi_0}$} &
\resizebox{8cm}{!}{\includegraphics{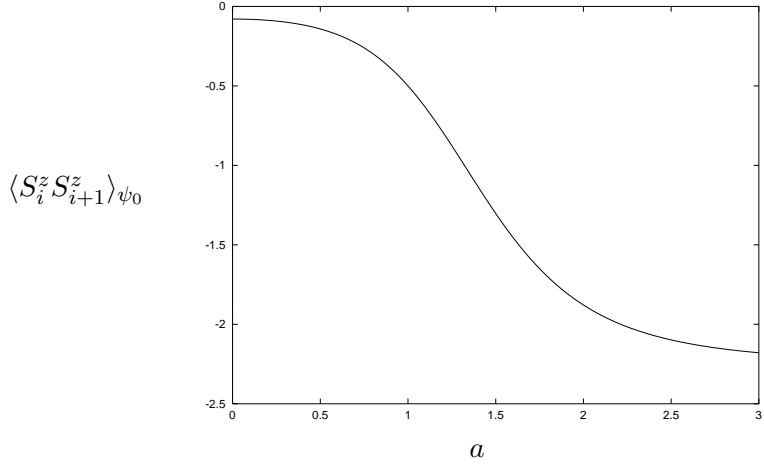}} \\
& $a$
\end{tabular}
\caption{\label{wafnnc}Nearest neighbour correlation of the weak antiferromagnet}
\end{figure}

The next interesting observable is the nearest neighbour correlation
\ras{1.5}
\be \begin{array}{l}
\langle S_i^z S_{i+1}^z\rangle_{\psi_0} \\
= -\langle S_m^z\rangle_{\psi_0}^2 - \frac{1}{2}\cdot\frac{(a^2 + 1)^2}
   {(a^2 - 1)^2 + 4}\cdot\frac{1-a^2+\sqrt{(a^2-1)^2 + 4}}
   {2 + \sqrt{(a^2 - 1)^2 + 4}} \quad ,
\end{array} \ee
\ras{1.0}
which is plotted as a function of the parameter $a$ in figure \ref{wafnnc}.
For large values of $a$, the nearest neighbour correlation approaches
$-\frac{9}{4}$, which is the value for a strong antiferromagnet with
fully polarized sublattices.

\begin{figure}[htb]
\begin{tabular}{rc}
\raisebox{3cm}{$\begin{array}{c} \xi_l^{-1} \\ \xi_t^{-1} \end{array}$} &
\resizebox{8cm}{!}{\includegraphics{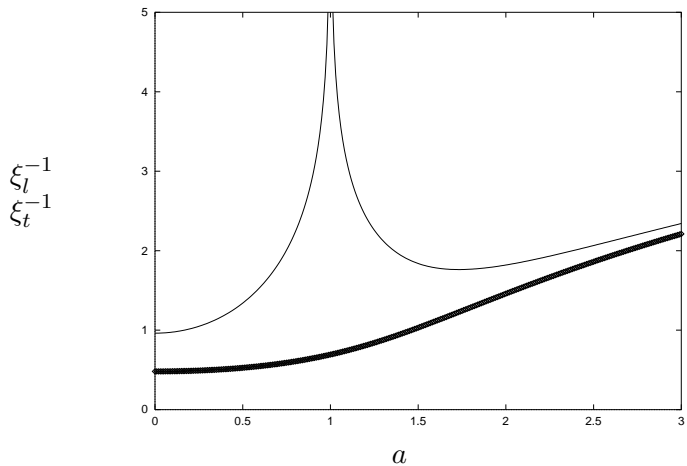}} \\
& $a$
\end{tabular}
\caption{\label{wafcl}Inverse longitudinal (thin) and transverse (thick) correlation length
                      of the weak antiferromagnet}
\end{figure}

The longitudinal correlation function consists of a constant and an
exponentially decaying term:
\be
\langle S_1^z S_r^z\rangle_{\psi_0} = \langle S_m^z\rangle_{\psi_0}^2 
   + A_l\, e^{-r/\xi_l} \quad\mbox{for odd }r\quad,
\ee
where $\xi_l$ is the longitudinal correlation length. Its inverse is given by
\be
\xi_l^{-1} = \ln\left|\frac{1+a^2+\sqrt{(a^2 - 1)^2 + 4}}
           {1+a^2-\sqrt{(a^2 - 1)^2 + 4}}\right|
\ee
and is plotted as a function of $a$ in figure \ref{wafcl}.
Since the inverse correlation length never vanishes, the system is non-critical. But at
$a=1$ it goes to infinity, i.~e.~the complete exponentially decaying term in the
correlation function vanishes, only the constant term survives. This term indicates
long-range antiferromagnetic order.

Unlike the longitudinal one, the transverse correlation function
$\langle S_1^x S_r^x\rangle_{\psi_0}$ does not exhibit a constant
term. It decays exponentially with inverse correlation length
\be
\xi_t^{-1} = \ln\frac{1+a^2+\sqrt{(a^2-1)^2+4}}{2} \quad ,
\ee
which is also plotted as a function of $a$ in figure \ref{wafcl}.
It never vanishes nor diverges.

Finally we note that in fact we have from (\ref{wafmpgstate}) two ground states
by exchanging sublattices, namely
\be \begin{array}{rcl}
|\psi_0^1\rangle & = & \tr m_1 \cdot g_2 \cdot m_3 \cdot g_4 \cdots g_L \quad\mbox{and} \\
|\psi_0^2\rangle & = & \tr g_1 \cdot m_2 \cdot g_3 \cdot m_4 \cdots m_L \quad .
\end{array} \ee
These are the only ground states of the global Hamiltonian $H$ for the closed chain.
The proof can be carried out using a straightforward induction over
the chain length $L$ and is omitted here. This degeneracy indicates that the sublattice
symmetry is broken, as is typical for a \neel\ antiferromagnet.
\subsection{Weak ferromagnet}
\label{wfgs}
Instead of using the alternating product of the matrices $m$ and $g$ defined in
equation (\ref{mandg}), it is also possible to construct global ground states
from the homogeneous products
\be
\label{wfmpgstate}
\begin{array}{rcl}
|\phi_0^+\rangle & = & \tr m_1 \cdot m_2 \cdot m_3 \cdots m_L \\
|\phi_0^-\rangle & = & \tr g_1 \cdot g_2 \cdot g_3 \cdots g_L \quad .
\end{array}
\ee
These two states are optimum ground states of $H$, if the local Hamiltonian $h$
annihilates the elements of $m\cdot m$ and $g\cdot g$, which are multiples of the
following 7 pair states:
\be \begin{array}{rcl}
|\zp{3}\zp{1}\rangle & + & \sigma |\zp{1}\zp{3}\rangle \\
|\zn{3}\zn{1}\rangle & + & \sigma |\zn{1}\zn{3}\rangle \\
|\zp{1}\zp{1}\rangle & + & a |\zp{3}\zn{1}\rangle \\
|\zp{1}\zp{1}\rangle & + & a |\zn{1}\zp{3}\rangle \\
|\zn{1}\zn{1}\rangle & + & a |\zn{3}\zp{1}\rangle \\
|\zn{1}\zn{1}\rangle & + & a |\zp{1}\zn{3}\rangle \\
|\zp{1}\zn{1}\rangle & + & \sigma |\zn{1}\zp{1}\rangle \quad .
\end{array} \ee
Therefore, if the parameters defined in equation~(\ref{genham}) obey
\be \begin{array}{l}
\lambda_2^\sigma    = \lambda_{11}^+      = \lambda_1^-            =
\lambda_{01}^\sigma = 0 \\
b = 0 \\
\lambda_3,\;\lambda_2^{-\sigma},\;\lambda_{12}^+,\;
\lambda_{01}^{-\sigma},\;\lambda_{02}^\sigma,\;\lambda_{02}^{-\sigma} > 0 \\
a,\;c \; \mbox{arbitrary,}
\end{array} \ee
then $|\phi_0^+\rangle$ and $|\phi_0^-\rangle$ are optimum ground states of the
global Hamiltonian~$H$. The system still contains 8 continuous parameters
(6 $\lambda$-parameters and $a$,$c$), which are not fixed. The global ground state,
however, only depends on one continuous ($a$) and one discrete parameter ($\sigma$).

As for the weak antiferromagnet, expectation values of local observables can be
calculated using the transfer matrix technique. The eigenvalues and the corresponding
eigenvectors can be found in Appendix \ref{appwftm}. It is sufficient to
calculate expectation values of $|\phi_0^+\rangle$, since those of
$|\phi_0^-\rangle$ are obtained by replacing $S^z$ by $-S^z$.
The magnetization per spin is exactly
\be
\langle S_i^z\rangle_{\phi_0^+} = \foh \quad ,
\ee
independent of the parameter $a$, which indicates a state of ferromagnetic ordering. But
unlike the strong spin-\fth\ ferromagnet, which has a magnetization of \fth\ per spin,
the system is not fully polarized. We therefore call this state a {\em weak ferromagnet}.

The fluctuations of $S_i^z$ are $a$-dependent and given by
\be
\langle \left(S_i^z\right)^2\rangle_{\phi_0^+} = \frac{1}{4} + \frac{|a|}{1+|a|}
\in \left[ \frac{1}{4},\frac{5}{4} \right[ \quad .
\ee
Because of rotational invariance in the $xy$-plane, this also yields
\be
\langle \left(S_i^x\right)^2\rangle_{\phi_0^+} =
\langle \left(S_i^y\right)^2\rangle_{\phi_0^+} =
\frac{7}{4} - \frac{|a|}{2(1+|a|)} \quad.
\ee

\begin{figure}[htb]
\begin{tabular}{rc}
\raisebox{3cm}{$\begin{array}{c} \xi_l^{-1} \\ \xi_t^{-1} \end{array}$} &
\resizebox{8cm}{!}{\includegraphics{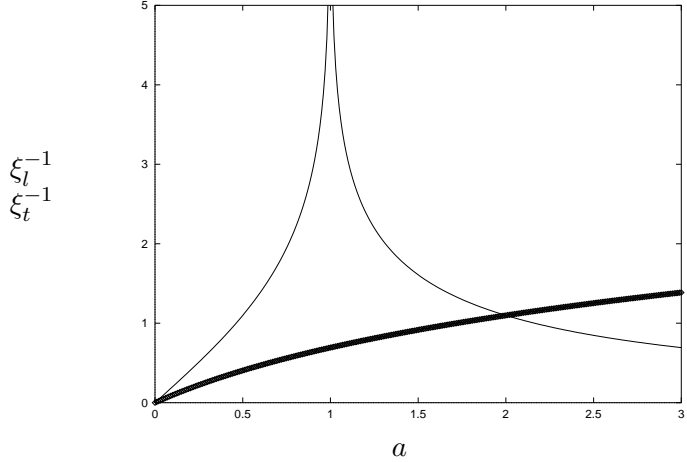}} \\
& $a$
\end{tabular}
\caption{\label{wfcl}Inverse longitudinal (thin) and transverse (thick) correlation length
                     of the weak ferromagnet}
\end{figure}

\begin{figure}[htb]
\begin{tabular}{rc}
\raisebox{3cm}{$\langle S_i^z S_{i+1}^z\rangle_{\phi_0^+}$} &
\resizebox{8cm}{!}{\includegraphics{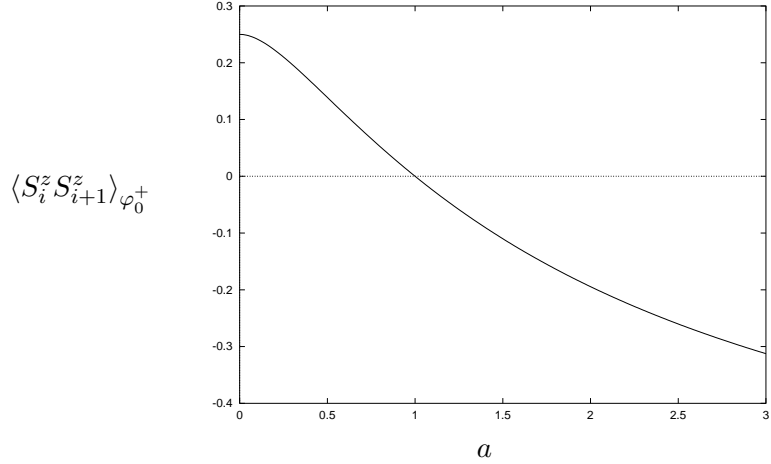}} \\
& $a$
\end{tabular}
\caption{\label{wfnnc}Nearest neighbour correlation of the weak ferromagnet}
\end{figure}

The longitudinal two-spin correlation function is
\be
\langle S_1^z S_r^z\rangle_{\phi_0^+} =
\frac{1}{4} - \frac{a^2}{(1-|a|)^2}\left(\frac{1-|a|}{1+|a|}\right)^r
\quad\mbox{for}\quad r\geq 2\quad,
\ee
which consists of a constant term -- indicating the long-range ferromagnetic order --
and an exponentially decaying term with inverse correlation length
\be
\xi_l^{-1} = \ln \left| \frac{1+|a|}{1-|a|} \right| \quad.
\ee
It is plotted as a function of $a$ in figure \ref{wfcl}.
Note the divergence at $a=1$, which also occured for the weak antiferromagnet and indicates
complete absence of truncated correlations for $r\geq 3$.
As special case we also get the nearest neighbour correlation
\be
\langle S_i^z S_{i+1}^z \rangle_{\phi_0^+} = \frac{1}{4} - \frac{a^2}{(1+|a|)^2} \quad.
\ee
Its $a$-dependence is shown in figure \ref{wfnnc}.
Although the global state has ferromagnetic long range order, the nearest neighbour
coupling is negative for $|a|>1$. This is due to the fact, that for large $|a|$ the
two states with $S^z=\fth$ and $S^z=-\foh$ are dominating in the $m$-matrix (\ref{mandg})
which leads to an alternation of these dominating states in the global state (\ref{wfmpgstate}).

Unlike the longitudinal one, the transverse two-spin correlation function decays
purely exponentially. The corresponding inverse correlation length
\be
\xi_t^{-1} = \ln (1+|a|)
\ee
is plotted as a function of $a$ in figure \ref{wfcl} together with the longitudinal one.

One can prove by induction with respect to $L=2,3,4,\ldots$, that $|\phi_0^+\rangle$ and
$|\phi_0^-\rangle$ are the only ground states of the global Hamiltonian $H$ for the closed chain.
The proof is completely analogous to the one for the weak antiferromagnet and is again omitted.
\subsection{Dimerized antiferromagnet}
\label{dafgs}
Unlike the weak antiferromagnet and the weak ferromagnet discussed before, a last type of optimum
ground states is not composed of square matrices. Instead, the alternating product of
the $2\times 3\,/\,3\times 2$ matrices,
\be
G = \left( \begin{array}{rrr}
         |\zn{1}\rangle & \sqrt{2}|\zp{1}\rangle &      a |\zp{3}\rangle \\
\sigma a |\zn{3}\rangle & \sqrt{2}|\zn{1}\rangle & \sigma |\zp{1}\rangle
           \end{array} \right) \quad
M = \left( \begin{array}{rr}
               |\zp{1}\rangle & a       |\zp{3}\rangle \\
\sqrt{2}\sigma |\zn{1}\rangle & \sqrt{2}|\zp{1}\rangle \\
             a |\zn{3}\rangle &         |\zn{1}\rangle
           \end{array} \right)
\ee
can be used to construct the global ground state
\be
\label{dafmpgstate}
|\Psi_0\rangle = \tr G_1 \cdot M_2 \cdot G_3 \cdot M_4 \cdots M_L \quad .
\ee
The elements of the products $G\cdot M$ and $M\cdot G$ are linear combinations of
the following 9 pair states:
\be \begin{array}{rcl}
\label{daflgs}
|\zp{3}\zp{1}\rangle & + & \sigma |\zp{1}\zp{3}\rangle \\
|\zn{3}\zn{1}\rangle & + & \sigma |\zn{1}\zn{3}\rangle \\
|\zp{1}\zp{1}\rangle & + & a |\zp{3}\zn{1}\rangle \\
|\zp{1}\zp{1}\rangle & + & a |\zn{1}\zp{3}\rangle \\
|\zn{1}\zn{1}\rangle & + & a |\zn{3}\zp{1}\rangle \\
|\zn{1}\zn{1}\rangle & + & a |\zp{1}\zn{3}\rangle \\
|\zp{1}\zn{1}\rangle & + & \sigma a^2 |\zp{3}\zn{3}\rangle \\
|\zn{1}\zp{1}\rangle & + & \sigma a^2 |\zn{3}\zp{3}\rangle \\
|\zp{1}\zn{1}\rangle & + & \sigma |\zn{1}\zp{1}\rangle \quad .
\end{array} \ee
These states are annihilated by the local Hamiltonian $h$ if the parameters in
equation~(\ref{genham}) satisfy
\be \begin{array}{l}
\label{dafpara}
\lambda_2^\sigma    = \lambda_{11}^+      = \lambda_1^-            =
\lambda_{01}^\sigma = \lambda_{02}^\sigma = \lambda_{01}^{-\sigma} = 0 \\
b = c = \sigma a^2 \\
\lambda_3,\;\lambda_2^{-\sigma},\;\lambda_{12}^+,\;\lambda_{02}^{-\sigma} > 0 \\
a \; \mbox{arbitrary,} \; \sigma=\pm 1 \quad .
\end{array} \ee
In this case, the local Hamiltonian has 9 ground states and 7 excited states. The
system still contains 5 continuous and 1 discrete parameter ($\sigma$). In the
isotropic case, i.~e.~$a=-\sqrt{3}$ and $\sigma=-1$, two spin correlation functions
have already been calculated in \cite{fmh1,fmh2}.

The transfer matrix and its eigensystem can be found in Appendix \ref{appdaftm}.
Unlike the ground states presented in subsections \ref{wafgs} and \ref{wfgs}, this state
is antiferromagnetic in the sense that $\langle S^z_{\mbox{\scriptsize total}}\rangle = 0$
and has vanishing sublattice magnetization
\be
\langle S_G^z \rangle_{\Psi_0} = \langle S_M^z \rangle_{\Psi_0} = 0 \quad .
\ee
As, however, the alternating $G$ and $M$ matrices are different, sublattice symmetry is broken.
The global ground state is therefore a dimerized antiferromagnet.

\begin{figure}[htb]
\begin{tabular}{rc}
\raisebox{3cm}{$\begin{array}{cc}\langle S_G^z S_M^z\rangle_{\Psi_0} \\
                                 \langle S_M^z S_G^z\rangle_{\Psi_0}\end{array}$} &
\resizebox{8cm}{!}{\includegraphics{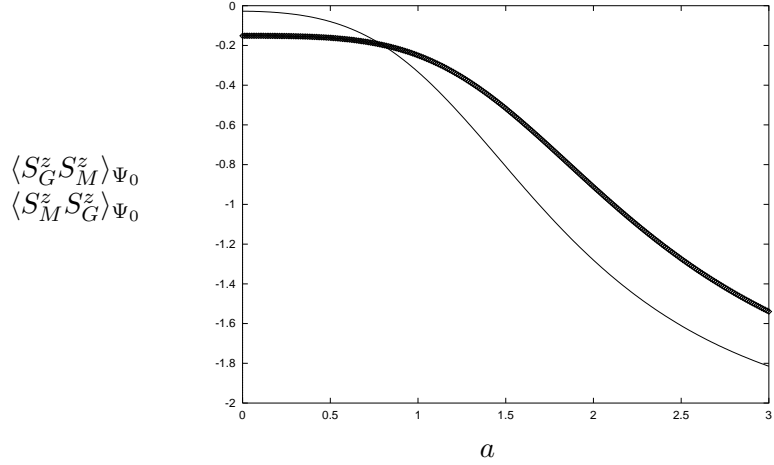}} \\
& $a$
\end{tabular}
\caption{\label{dafnnc}$G\cdot M$ (thin) and $M\cdot G$ (thick) nearest neighbour correlation
                       of the dimerized antiferromagnet}
\end{figure}

The fluctuations of $S^z$ are given by
\be
\langle (S_G^z)^2\rangle_{\Psi_0} = \langle (S_M^z)^2\rangle_{\Psi_0} =
   \frac{9}{4} - \frac{2 a^2 + 18}{(a^2 + 1)^2 + 8} \in
   \left[\frac{1}{4},\frac{9}{4}\right[ \quad ,
\ee
thus covering the complete range of possible values. Because of rotational invariance we
also have
\be
\langle (S^x)^2\rangle_{\Psi_0} = \langle (S^y)^2\rangle_{\Psi_0} =
   \frac{3}{4} + \frac{a^2 + 9}{(a^2 + 1)^2 + 8} \quad .
\ee
As $G$ and $M$ are not square matrices, there are {\em two different} nearest
neighbour correlations in contrast to the situation in subsections \ref{wafgs} and \ref{wfgs}.
If the left matrix is of $G$ type, we have
\be
\langle S_G^z S_M^z\rangle_{\Psi_0} = 
   -\frac{9}{4}\cdot\frac{(a^2 + \frac{1}{3})^2}{(a^2 + 1)^2 + 8} \quad ,
\ee
otherwise it is
\be
\langle S_M^z S_G^z\rangle_{\Psi_0} =
   -\frac{9}{4}\,\left[\,\frac{(a^2 + \frac{1}{3})^2 + \frac{20}{9}}
                              {(a^2 + 1)^2 + 8}\,\right]^2 \quad .
\ee
Both are plotted as a function of $a$ in figure \ref{dafnnc}.
Except for one special value of $a$, these two correlations do not coincide.
The alternating binding structure characterizes the broken symmetry.
This type of state is called {\em dimerized}.

\begin{figure}[htb]
\begin{tabular}{rc}
\raisebox{3cm}{$\begin{array}{c} \xi_l^{-1} \\ \xi_t^{-1} \end{array}$} &
\resizebox{8cm}{!}{\includegraphics{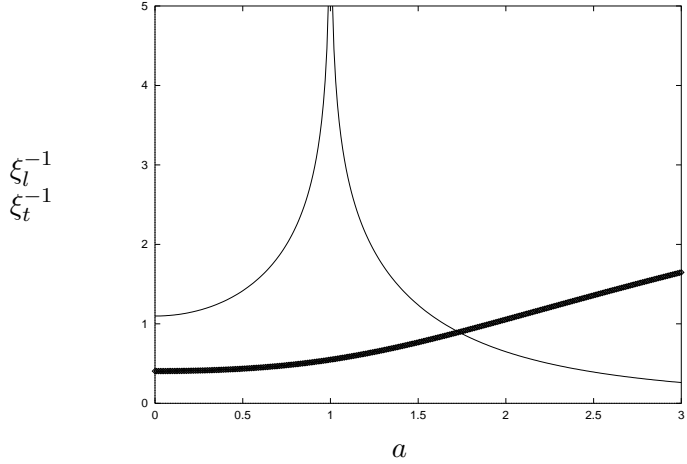}} \\
& $a$
\end{tabular}
\caption{\label{dafcl}Inverse longitudinal (thin) and transverse (thick) correlation length
                      of the dimerized antiferromagnet}
\end{figure}

Both the longitudinal and the transverse correlation functions decay exponentially
\be \begin{array}{rcl}
\langle S_1^z S_r^z\rangle_{\Psi_0} & = & A_l \, e^{-r/\xi_l}
  \hspace{1cm}\mbox{for odd values of $r$} \\
\langle S_1^x S_r^x\rangle_{\Psi_0} & = & A_t \, e^{-r/\xi_t} \quad .
\end{array} \ee
The corresponding inverse correlation lengths
\be \begin{array}{rcl}
\xi_l^{-1} & = & \ln \frac{\sqrt{(a^2 + 1)^2 + 8}}{|a^2 - 1|} \\
\xi_t^{-1} & = & \ln \frac{1}{2}\sqrt{(a^2 + 1)^2 + 8}
\end{array} \ee
are plotted as a function of $a$ in figure \ref{dafcl}.
As both do not vanish for any finite value of $a$, there is no critical behaviour.
At $a=1$ the longitudinal correlation length again diverges as in subsections
\ref{wafgs} and \ref{wfgs}.

The list of local ground states (\ref{daflgs}) actually is the combination of all local
ground states of the weak antiferromagnet (section \ref{wafgs}) and the weak ferromagnet
(section \ref{wfgs}). Therefore it is clear that the corresponding global states are also optimum
ground states of the global Hamiltonian with parameters (\ref{dafpara}). However, it turns out
that the degeneracy is even much higher. {\em Any} state of the form
\be
\label{dimdeg}
|\Psi\rangle = \tr p_1\cdot p_2\cdot p_3\cdots p_L
\ee
is an optimum ground state of the global Hamiltonian $H$, if every single matrix
$p_i$ is one of the matrices $m$ or $g$ (\ref{mandg}) from subsection \ref{wafgs}.
The reason for this is
that the local Hamiltonian $h$ annihilates the elements of {\em all} four possible products
of $m$ and $g$:
\be
h(m\cdot m) = h(g\cdot g) = h(m\cdot g) = h(g\cdot m) = 0 \quad .
\ee
We therefore have a global degeneracy of at least $n_0 = 2^L$. Among these states are the following
homogeneous special cases. The two states
\be \begin{array}{rcl}
|\phi_0^+\rangle & = & \tr m_1 \cdot m_2 \cdot m_3 \cdots m_L \\
|\phi_0^-\rangle & = & \tr g_1 \cdot g_2 \cdot g_3 \cdots g_L
\end{array} \ee
represent the weak ferromagnet discussed in subsection \ref{wfgs}. The two states
\be \begin{array}{rcl}
|\psi_0^1\rangle & = & \tr m_1 \cdot g_2 \cdot m_3 \cdot g_4 \cdots g_L \\
|\psi_0^2\rangle & = & \tr g_1 \cdot m_2 \cdot g_3 \cdot m_4 \cdots m_L
\end{array} \ee
represent the weak antiferromagnet discussed in subsection \ref{wafgs}.
Finally, we have the dimerized antiferromagnet (\ref{dafmpgstate}) of this subsection.
\section{Vertex state model representation}
\label{vertrep}
Apart from the foregoing matrix representation which is especially convenient for
calculational purposes,
it is also possible to give a graphic representation of the constructed optimum ground
states in terms of vertices. This {\em vertex state model representation} is especially
useful to construct optimum ground states in higher dimensions \cite{nkz}.

For a specific matrix $m$ we map each element on a vertex with two emanating bonds.
A variable is attached
to the left and the right bond corresponding to the left and right matrix index, respectively.
The bond variables are represented by arrows pointing out of or into the vertex:
\be \begin{array}{lcl}
m_{11} & \mapsto & \vertcll \\
m_{12} & \mapsto & \vertclr \\
m_{21} & \mapsto & \vertcrl \\
m_{22} & \mapsto & \vertcrr \quad .
\end{array} \ee
The notation for a matrix product is
\be \begin{array}{rcl}
(m\cdot m)_{12} & \mapsto & \vvertclra + \vvertclrb \\
                & =       & \vvertclr \quad,
\end{array} \ee
i.~e.~a contraction denotes the sum over the bond variable.
Since we are interested in matrices, whose elements are single spin states, we must associate
a weighted spin state with each vertex rather than a number (vertex weight) in the usual case.
The result will therefore be a ``vertex state model'' rather than a ``vertex model''.

In order to describe global ground states composed of more than one type of
matrices, it is also necessary to introduce different classes of vertices.
This leads us to the following classification of vertex states corresponding to (\ref{mandg}):
\be \begin{array}{rcrl}
\vertcll & : &          |\zp{1}\rangle & \mbox{($m$-class)} \\
\vertclr & : &        a |\zp{3}\rangle & \\
\vertcrl & : &          |\zn{1}\rangle & \\
\vertcrr & : &   \sigma |\zp{1}\rangle & \\
         &   &                         & \\
\vertoll & : &   \sigma |\zn{1}\rangle & \mbox{($g$-class)} \\
\vertolr & : &          |\zp{1}\rangle & \\
\vertorl & : &        a |\zn{3}\rangle & \\
\vertorr & : &          |\zn{1}\rangle & \quad .
\end{array} \ee
As an example the following graph represents the ground state $|\psi_0\rangle$ of the
weak antiferromagnet from subsection \ref{wafgs} for the closed chain with $L=8$.
\be
|\psi_0\rangle \; \Longrightarrow \vertring \quad .
\ee
We note that for a given class of vertices, the $S^z$ eigenvalue of the vertex state
is given by
\be \begin{array}{rcl}
S^z & = & \foh (\mbox{\# of outgoing arrows}  -
                \mbox{\# of ingoing  arrows}) \\
& + & \left\lbrace
\begin{array}{rcl}
 \foh & , & m-\mbox{class} \\
-\foh & , & g-\mbox{class}
\end{array} \right. \quad .
\end{array}
\ee
This also ensures conservation of the local pair magnetization $S_i^z + S_{i+1}^z$.

The $2\times 3$- and $3\times 2$-matrices from subsection \ref{dafgs} can also be handled
easily using
a new type of bond that carries a bond variable with values: `left arrow', `zero', and
`right arrow'.
\section{Summary}
\label{summary}
We have presented three different one-parameter sets of optimum ground states for the spin-\fth\
chain. The first one is a \neel\ ordered antiferromagnet which exhibits a parameter dependent
sublattice magnetization. Full polarization of the sublattices is only achieved in the limit
$a\to\infty$. We therefore call this state a {\em weak antiferromagnet}. It is an optimum ground
state for a 6-dimensional manifold of Hamiltonians with usual symmetries. The global degeneracy
is 2. Truncated correlations decay exponentially.

The second set of states is of ferromagnetic type. The global magnetization is $\frac{L}{2}$,
which is one third of the maximum value. So this is a {\em weak ferromagnet}. The corresponding
set of Hamiltonians is an 8-dimensional manifold. Again the global degeneracy is 2, corresponding
to the broken $S^z \rightarrow -S^z$ symmetry. Also truncated correlations decay exponentially.

Finally we have the {\em dimerized antiferromagnet}. Its properties are quite different from the
first two types of ground states. There is no sublattice magnetization, but the broken sublattice
symmetry here leads to an alternating binding structure of nearest neighbour spins. The
corresponding set of Hamiltonians is a 5-dimensional manifold, thus smaller than in the two other
cases. Actually we have a vast number of ground states of the type (\ref{dimdeg}), the degeneracy is at
least $2^L$, corresponding to a specific zero-temperature entropy of ln 2.
\appendix
\setcounter{equation}{0}
\renewcommand{\theequation}{\Alph{section}.\arabic{equation}}
\renewcommand{\thesection}{Appendix \Alph{section}:}
\renewcommand{\thesubsection}{\Alph{section}.\arabic{subsection}}
\section{Transfer matrices}
The transfer matrix method has been explained in detail in \cite{ksz1}. In the following, we only
cite the main steps of this technique and summarize the results for the
optimum ground states presented in section \ref{gs32chain}.

To calculate the expectation value $\langle A_i \rangle$
of a local operator $A_i$ acting on spin states at site $i$ within the state (\ref{mpgstate}), we
define the transfer matrix
\be
\label{transmat}
T_{(k_1,l_1),(k_2,l_2)} = \langle \,(m\cdot g)_{k_1,k_2} \,| \,(m\cdot g)_{l_1,l_2}\,\rangle \quad .
\ee
Redefining index pairs as
\be \begin{array}{rcl}
\label{indexmap}
(1,1) & \longmapsto & 1 \\
(2,2) & \longmapsto & 2 \\
(1,2) & \longmapsto & 3 \\
(2,1) & \longmapsto & 4
    \end{array} \ee
the tensor (\ref{transmat}) results in a $4\times 4$-matrix $T$. Additionally, for any operator $A$
acting on $(m\cdot g)$ we define a tensor
\be
T_{(k_1,l_1),(k_2,l_2)}(A) = \langle \,(m\cdot g)_{k_1,k_2} \,|A| \,(m\cdot g)_{l_1,l_2}\,\rangle \quad ,
\ee
which also reduces to a $4\times 4$-matrix $T(A)$.
These two matrices can now be used to calculate the expectation value for a single spin
operator $A_i$ as
\be
\label{expect}
\langle A_i \rangle \equiv \frac{\langle\Psi_0|A_i|\Psi_0\rangle}{\langle\Psi_0|\Psi_0\rangle}
= \frac{\mbox{trace } T(A) T^{L/2-1}}{\mbox{trace } T^{L/2}} \quad .
\ee
Usually, $T$ is a symmetric matrix, so we can rewrite (\ref{expect}) in terms of the
eigenvalues $\chi_n$ and eigenvectors $u_n$ of $T$:
\be
\langle A_i \rangle =
\frac{\sum_{n=1}^4 \,\langle u_n|T(A)|u_n \rangle \; \chi_n^{L/2-1}}{\sum_{n=1}^4 \chi_n^{L/2}}
\quad .
\ee
Similar considerations also yield an expression for correlation functions. If $A_i$ and $B_j$
are operators acting on spins at sites $i$ and $j$, respectively,
we get
\be
\langle A_1 B_r \rangle =
\frac{\sum_{n_1,n_2=1}^4 \,\langle u_{n_1}|T(A)|u_{n_2} \rangle \; \chi_{n_2}^{r-2} \; 
                           \langle u_{n_2}|T(B)|u_{n_1} \rangle \;
 \chi_{n_1}^{L/2-r}}{\sum_{n=1}^4 \chi_n^{L/2}}
\ee
as the final result for all correlation functions of two local operators. Because of translational
invariance, we fix the first site at $i=1$ and denote the second site $j=r$, $r=2,3,\ldots\;$.
\subsection{Weak antiferromagnet}
\label{appwaftm}
For the ground state $|\psi_0\rangle$ (\ref{wafmpgstate}), the transfer matrix (\ref{transmat})
is given by
\be
T_{\psi_0} = \left(\begin{array}{cccc}
1+a^4 & 1+a^2 & 0 & 0 \\
1+a^2 & 2     & 0 & 0 \\
0     & 0     & 1 & 0 \\
0     & 0     & 0 & 1
\end{array} \right) \quad .
\ee
The eigenvalues $\chi_n$ and the corresponding eigenvectors $u_n$ are
\be \begin{array}{lcl}
\chi_1 = (1+x+\sqrt{1+x^2})^2 & , & u_1 = \frac{1}{\gamma_1} (1,\beta_1,0,0) \\
\chi_2 = (1+x-\sqrt{1+x^2})^2 & , & u_2 = \frac{1}{\gamma_2} (1,\beta_2,0,0) \\
\chi_3 = 1                    & , & u_3 = (0,0,1,0) \\
\chi_4 = 1                    & , & u_4 = (0,0,0,1) \quad ,
\end{array} \ee
where
\be \begin{array}{lclclcl}
x & = & \frac{1}{2}\left( a^2 - 1 \right) & & & & \\
\beta_1 & = & - x + \sqrt{1+x^2} & , & \gamma_1^2 & = & 1 + \beta_1^2 \\
\beta_2 & = & - x - \sqrt{1+x^2} & , & \gamma_2^2 & = & 1 + \beta_2^2 \quad .
\end{array} \ee
\subsection{Weak ferromagnet}
\label{appwftm}
In order to get a symmetric transfer matrix, we insert an identity matrix between
the single state matrices $m$ in (\ref{wfmpgstate}):
\be
\left(\begin{array}{cc}          \sqrt{|a|} &0\\0&1 \end{array} \right) \cdot
\left(\begin{array}{cc} \frac{1}{\sqrt{|a|}}&0\\0&1 \end{array} \right) = 1_{2\times 2}
\quad .
\ee
This leads us to a modified single state matrix
\be
\tilde{m} = \left(\begin{array}{rr} |\zp{1}\rangle & \mbox{sgn } a \sqrt{|a|}\,|\zp{3}\rangle \\
                        \sqrt{|a|}\,|\zn{1}\rangle & \sigma                  \,|\zp{1}\rangle
                  \end{array} \right) \quad ,
\ee
which obviously produces the same global state (\ref{wfmpgstate}) as $m$. The transfer matrix for
$\tilde{m}$ is
\be
T_{\phi_0^+} = \left( \begin{array}{cccc} 1&|a|&0&0\\|a|&1&0&0\\0&0&\sigma&0\\0&0&0&\sigma
                      \end{array} \right) \quad .
\ee
Its eigensystem is
\be \begin{array}{lcl}
\chi_1 = 1+|a|  & , & u_1 = \frac{1}{\sqrt{2}} (1, 1,0,0) \\
\chi_2 = 1-|a|  & , & u_2 = \frac{1}{\sqrt{2}} (1,-1,0,0) \\
\chi_3 = \sigma & , & u_3 = (0,0,1,0) \\
\chi_4 = \sigma & , & u_4 = (0,0,0,1) \quad .
\end{array} \ee
\subsection{Dimerized antiferromagnet}
\label{appdaftm}
As $G$ and $M$ alternate in the global state (\ref{dafmpgstate}), one can use the transfer matrix
for $M\cdot G$ or for $G\cdot M$. However,
since the transfer matrix for $M\cdot G$ would be a $9\times 9$-matrix, it is more economical
to use the transfer matrix for $G\cdot M$, which is a symmetric $4\times 4$-matrix, namely
\be
T_{\Psi_0} = \left( \begin{array}{cccc}
  a^4 + 5 & 2 a^2 + 4 & 0 & 0 \\
2 a^2 + 4 &   a^4 + 5 & 0 & 0 \\
        0 &         0 & 4 & 0 \\
        0 &         0 & 0 & 4
              \end{array} \right) \quad .
\ee
It has the following eigenvalues and eigenvectors:
\be \begin{array}{lcl}
\chi_1 = (a^2 + 1)^2 + 8 & , & u_1 = \frac{1}{\sqrt{2}} (1, 1,0,0) \\
\chi_2 = (a^2 - 1)^2     & , & u_2 = \frac{1}{\sqrt{2}} (1,-1,0,0) \\
\chi_3 = 4               & , & u_3 = (0,0,1,0) \\
\chi_4 = 4               & , & u_4 = (0,0,0,1) \quad .
\end{array} \ee

\end{document}